\documentclass{article}

\usepackage{arxiv}

\usepackage[utf8]{inputenc} 
\usepackage[T1]{fontenc}    
\usepackage{hyperref}       
\usepackage{url}            
\usepackage{booktabs}       
\usepackage{amsfonts}       
\usepackage{nicefrac}       
\usepackage{microtype}      
\usepackage{lipsum}		
\usepackage{graphicx}
\usepackage{natbib}
\usepackage{doi}

\title{A Tool for Test Case Scenarios Generation Using Large Language Models}

\author{ \href{https://orcid.org/0000-0000-0000-0000}{\includegraphics[scale=0.06]{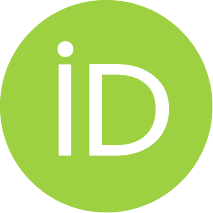}\hspace{1mm}Abdul Malik Sami}\\
        Tampere University\\
	\texttt{malik.sami@tuni.fi} \\
	\And
	\href{https://orcid.org/0000-0000-0000-0000}{\includegraphics[scale=0.06]{orcid.pdf}\hspace{1mm}Zeeshan Rasheed} \\
	Tampere University\\
	\texttt{zeeshan.rasheed@tuni.fi} \\
	\AND
	Muhammad Waseem \\
	Jyväskylä University \\
	\texttt{muhammad.m.waseem@jyu.fi} \\
	\And
	Zheying Zhang \\
	Tampere University \\
	\texttt{zheying.zhang@tuni.fi} \\
        \And
	Herda Tomas \\
	Austrian Postal Service \\
	\texttt{herda.tom@gmail.com} \\
        \And
	Pekka Abrahamsson \\
	Tampere University \\
	\texttt{pekka.abrahamsson@tuni.fi} \\
}

\renewcommand{\shorttitle}{\textit{arXiv} Template}

\hypersetup{
pdftitle={A template for the arxiv style},
pdfsubject={q-bio.NC, q-bio.QM},
pdfauthor={David S.~Hippocampus, Elias D.~Striatum},
pdfkeywords={First keyword, Second keyword, More},
}

\begin{document}
\maketitle

\begin{abstract}
Large Language Models (LLMs) are widely used in Software Engineering (SE) for various tasks, including generating code, designing and documenting software, adding code comments, reviewing code, and writing test scripts. However, creating test scripts or automating test cases demands test suite documentation that comprehensively covers functional requirements. Such documentation must enable thorough testing within a constrained scope and timeframe, particularly as requirements and user demands evolve. This article centers on generating user requirements as epics and high-level user stories and crafting test case scenarios based on these stories. It introduces a web-based software tool that employs an LLM-based agent and prompt engineering to automate the generation of test case scenarios against user requirements. 
\end{abstract}


\section{Introduction}
Requirements engineering, a critical discipline in software engineering, establishes the essential link between the technical specifications that software engineers develop and the intended purposes of the systems \cite{bjarnason2016multi}. Throughout the development process—spanning proposal development, design, implementation, and testing—requirements are continually discussed \cite{9487986}. These discussions ensure that the application meets user requirements, which may be detailed as user stories or tasks. A significant challenge in this process is accurately converting user requirements into technical specifications that precisely align with user needs and expectations.

Once the requirements are clearly defined, software testing plays a pivotal role in confirming that these requirements are met according to the acceptance criteria \cite{abrahamsson2017agile}. It underscores the importance of verifying requirement fulfillment and maintaining traceability to identify any user changes or discrepancies in development against the specified requirements \cite{mucha2024systematic}. In response to the increasing demands for rapid development, the field of software engineering is evolving at an accelerated pace, therefore it's critical to identify accurate, and clear requirements that are needed for using a test case scenarios suite using software requirements to effectivity validate and verify the requirements and its output \cite{tang2024chatgpt, svensson2024not}.

Artificial Intelligence (AI) and Natural Language Processing (NLP) initiate a revolutionary phase in computational tools, enabling the generation of text with human-like precision. The advent of advanced language models, especially within the GPT series, signifies a crucial development in this field \cite{radford2018improving,rasheed2024can, ouyang2022training}. These Large Language Models (LLMs) excel in understanding natural language, positioning themselves as pioneers in a significant shift within software engineering. Transitioning from simple tools to collaborative partners, they provide insights and analyses that were once beyond the reach of traditional approaches \cite{sami2024system, rasheed2023autonomous}. 
Using LLMs to convert requirements into user stories and then create test case scenarios is not only a theoretical undertaking but also a necessary step in the constantly changing software development industry. For each software's functional requirements, this progression satisfies the essential need for automating, enhancing, and perfecting requirement creation and validation through test case scenarios \cite{tikayat2023agile, zhang2024llm}.

\subsection{Problem Statement}

Despite the efforts to translate user requirements into detailed specifications for validation, the automation of test case scenarios for each functional requirement remains a notable challenge. The adoption of Large Language Model (LLM)-based agents to generate test case suites presents a promising avenue to overcome this obstacle.This study aims to explore the capabilities of LLMs in enhancing the generation of test case scenarios for functional requirements, with a focus on their application and impact within software requirement engineering and software testing perspectives. Our investigation is guided by the following research questions in Table 1:

\begin{table}[ht]
\centering
\caption{Exploring LLMs in Generating Test Case Scenarios from User Requirements}
\label{tab:llms_research_questions}
\begin{tabular}{|p{0.4\textwidth}|p{0.4\textwidth}|}
\hline
\textbf{Research Question} & \textbf{Objective} \\
\hline
How can large language models be applied to generating test case scenarios and suites? & We aim to understand how LLMs are utilized in software engineering, specifically for creating test case scenarios for each functional requirement. \\
\hline
What are the limitations and opportunities of using large language models for test suite generation and test case scenario creation? & Our goal is to identify the challenges and potential advancements in using LLMs for generating test suites and scenarios for software testing. \\
\hline
\end{tabular}
\end{table}

Our research strategy focuses on enhancing the existing tool \cite{sami2024system} through the application of Large Language Models (LLMs) in several key areas:
\begin{enumerate}
    \item Taking the input of existing work as an output of prioritizing these stories.
    \item Use OpenAI's agent-style API for creating test case scenarios with the aid of prompt engineering and LLMs.
    \item Allow the test suites to be downloaded in CSV format, enabling integration with a variety of test case management tools.
    \item Conduct assessments of performance and content analysis on the responses generated.
\end{enumerate}

In this paper, we report on the extension and enhancement of an existing web-based software tool designed for generating software test case scenarios. This work demonstrates the capabilities of GPT models in generating test scenarios from user stories. Our prototype highlights the potential of GPT models to interpret context from user stories and develop a comprehensive test suite effectively. This process assists testers in thoroughly verifying and validating requirements against every edge case and conceivable user scenario.

The following is how the paper is organized: Background study is presented in Section \ref{Background}, which establishes the context for our investigation. The research concept and methodology are described in Section \ref{Methodology}, which also provides an overview of our methodical approach. 
Section \ref{preliminary_results} presents our observations and examines the preliminary outcomes. The limitations of our study are discussed in Section \ref{challenges} along with suggestions for future research topics. Section \ref{conclusions} provides a conclusion.

\section{Background}
\label{Background}
\subsection{Generative AI Role in software testing}
\label{Generative AI}

Generative AI is designed to reproduce human-like output in a variety of fields, including multimedia production, computer vision, and natural language processing (NLP) \cite{hacker2023regulating}. The goal is to generate new material through this process. Through the use of autoregressive language models like GPT and Generative Adversarial Networks (GANs), this technology has transformed text production, machine translation, conversational systems, and code generation \cite{aydin2023chatgpt, waseem2023using}. The transformer architecture, which greatly improves NLP skills by recognizing contextual relationships in text, is essential to the GPT model's capacity to produce text that nearly resembles human writing \cite{radford2018improving,rasheed2024large}.

Recent advancements highlight how versatile and highly effective GPT models are, and how they can revolutionize software engineering (SE) methods \cite{liu2023gpt,ouyang2022training}. These models allow the automation of a number of tasks in the software development lifecycle, such as error identification, code snippet creation, and documentation, by using large code databases for training \cite{feng2023investigating,treude2023navigating}. Incorporating GPT models into SE improves software production efficiency and quality overall, while also accelerating coding and application development \cite{dong2023self,ma2023scope}.

The appearance of generative AI in SE marks a noteworthy shift, underscoring its capability to refine and expedite software development processes. Incorporating GPT models into SE heralds a pivotal transformation, affecting how software is crafted and maintained. Crucially, this advancement streamlines the prioritization of requirements—a key phase in development. AI's application enables developers to generate test suite and test case scenarios more effectively, ensuring that critical features align with user needs and organizational objectives. This approach promises a move towards a more user-centric, and efficient paradigm in software development and testing.

Many studies have looked into using GPT models in software engineering (SE) and software testing, including unit and GUI testing (\cite{ahmad2023towards, barke2023grounded}. 
Using Large Language Models (LLMs) for developing test suites is a big step forward that can make software development more efficient and effective. Yet, issues like false information (hallucinations) and the limitations of understanding natural language show there's a big need for research. 
We need a solution that can grasp wide-ranging client needs, automatically create user stories, and use LLMs to organize testing tasks efficiently.
Modern software development methods, which value flexibility and the smart use of LLM agents, should be adopted by this tool. 
By addressing this need, we can enhance software quality and make full use of AI in the software development lifecycle. 
This opens up great possibilities for new research and development in the field by integrating test suites as user stories are being created, ensuring all critical scenarios are considered and meet user needs.

\section{Preliminary Methodological Framework}
\label{Methodology}
We enhance our existing web-based tool in this research by employing LLMs to provide a user-centric solution for precisely and efficiently generating test case scenarios for software development requirements. This tool benefits from the integration of React, Flask, and OpenAI technologies.

\begin{enumerate}
    \item \textbf{Tool Development:} We enhance the capabilities of our existing tools to demonstrate the practical application of LLMs in generating test cases that meet user requirements.
    
    \item \textbf{Test Case Scenarios Generation:} Users can input their requirements or upload predefined user stories. The system utilizes LLMs to generate test cases and scenarios for each requirement, covering all aspects of the use case.
    
    \item \textbf{Evaluation and Output:} During the evaluation phase, we analyze the tool's outputs, including performance and content analysis generated by LLMs, and verify the conformance.
\end{enumerate}

Our methodology not only empirically validates the effectiveness of our tool but also illuminates the practical benefits of integrating LLMs into the software testing life cycle and the documentation process. It enhances the efficiency and ensures requirements align with user needs and modern software development practices.

\begin{figure*}[t]
    \centering
    \includegraphics[width=\linewidth]{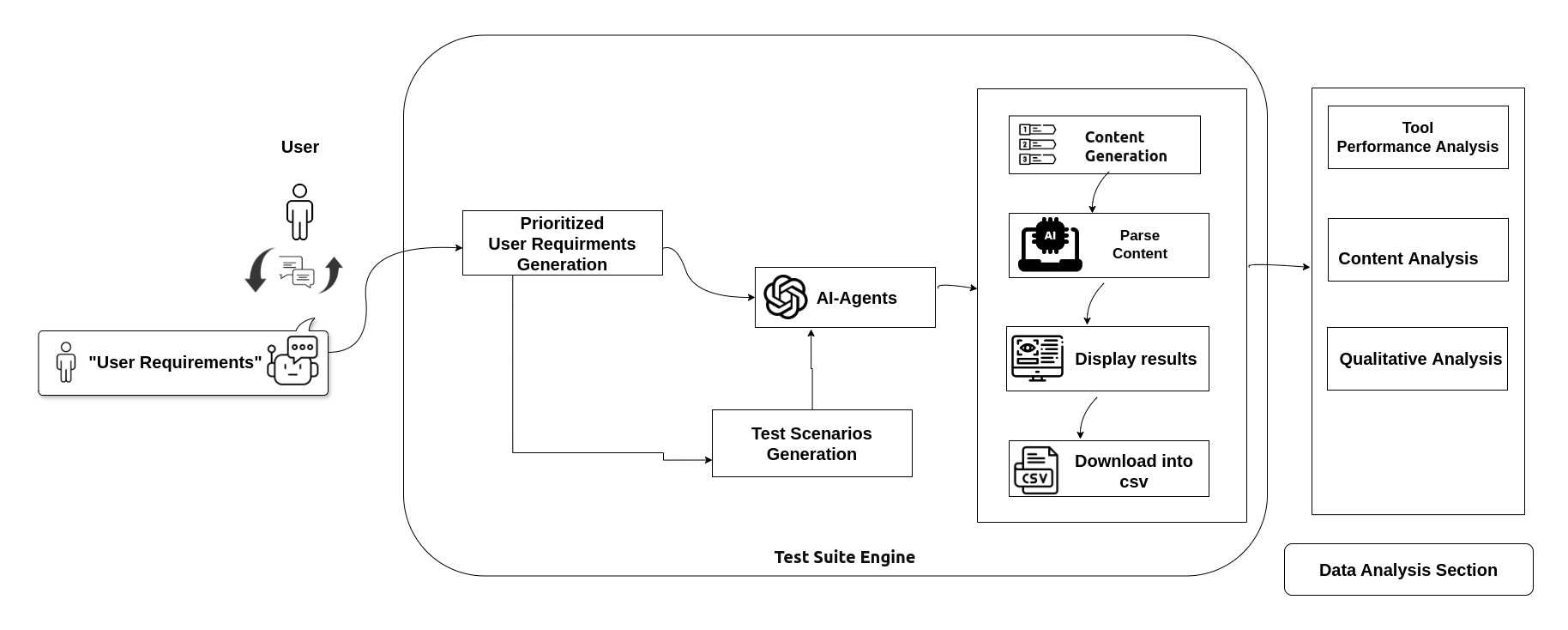}
    \caption{Process of Generating testCase Scenarios Using LLMs and Analyzing Its Output}
    \label{fig:result}
\end{figure*}

\section{Preliminary Empirical Results}
\label{preliminary_results}

Our research introduces a system that streamlines the generation of test case scenarios using LLM agent. The aim is to enhance research workflows utilizing a Python Flask API, React UI, and ChatGPT 3.5. 
Our initial evaluation focuses on creating test case scenarios for documentation purposes. We perform an early analysis to assess system effectiveness and pinpoint areas for enhancement based on user feedback.

\subsection{Identified User Requirements and Test Scenarios}
Our development is driven by specific user requirements.  we generate test case scenarios that are meticulously checked and tailored to meet user needs.
Fulfilling these requirements is crucial for maximizing research productivity and the efficiency of the process. Also noticable thing The quality of the input matters and test case generation change of the nature of user requirements.

\subsection{System Configuration}
The architecture of our system incorporates React, CSS, and JavaScript with ANT design for the frontend, complemented by Flask for backend operations. This setup provides an API that interacts with the GPT-3.5 model, which transforms user inputs into organized and prioritized user story-based test cases.

\subsection{Working of the tool}

\textbf{Step 1: Generation of Test case scenarios:} The system features functionality for generating test case scenarios againt the list in JSON format. 

\textbf{User Story:} As a researcher, I aim to formulate questions that align with my research objectives in order to direct the focus of the SLR-GPT system towards pertinent topics.

Within this context, several scenarios are envisaged. For instance, \textbf{Test Case 1} involves a researcher providing clear keywords related to their research objective, like \emph{``machine learning techniques for image recognition.''} The expectation is that the SLR-GPT system would generate questions honing in on the key aspects of machine learning for image recognition.

Another scenario, \textbf{Test Case 5}, presents a situation where a researcher sets forth objectives with specific constraints, such as \emph{``ethical considerations in AI applications under GDPR regulations.''} In this case, the anticipated output is that the SLR-GPT system would customize questions to concentrate on the ethical considerations within AI applications, specifically concerning GDPR compliance.

 The testcases and its scenarios are listed displayed in Figure 2.

\begin{figure*}[t]
    \centering
    \includegraphics[width=\linewidth]{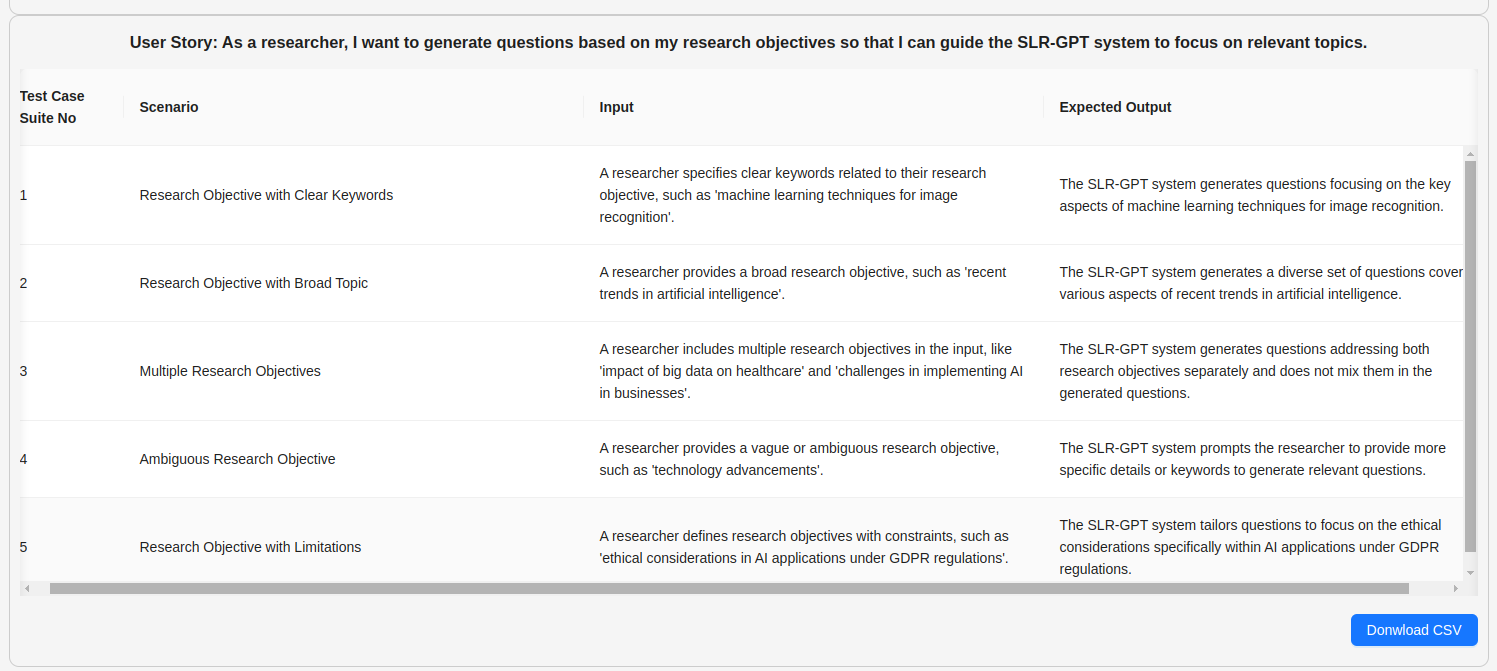}
    \caption{Test Scenarios generation Using LLMs}
    \label{fig:third_step_prioritization}
\end{figure*}

\textbf{Step 2: Exporting Test cases:} The system simplifies the exportation of test case scenarios into a CSV format, facilitating their integration with project management and testing tools like JIRA, Azure DevOps, and RE management tools.

The efficiency of this process is demonstrated by its ability to generate test case scenarios for each prioritized user requirement within an average of 2 seconds, highlighting the method's rapid execution and the system's capacity to accelerate the documentation of test case scenarios.

In performing content analysis, we are calling the agent API, which is using regular expressions to parse the natural language responses. Making adjustments to various parameters, like temperature, proves essential for achieving accuracy. By conducting several rounds of testing, we are consistently producing results that match the expected behavior.

\section{Challenges and Future Directions}
\label{challenges}
This study makes use of LLMs for generating documents aimed at creating test case scenarios, but we have identified several challenges. 

\begin{enumerate}
    \item \textbf{LLM Hallucinations:} A prevalent issue is LLM hallucinations, where the models generate outputs that do not correspond with the given input data \cite{xu2024hallucination}. Incorporating Retrieval-Augmented Generation (RAG) and other fine-tuning techniques may mitigate this problem.
    
    \item \textbf{Content Analysis Automation:} It is essential to automate the process of content analysis and to ensure its accuracy. Our preliminary study provides a proof of concept that demonstrates LLMs' abilities to produce test suites and edge case scenarios from requirements, yet there is a necessity to further refine and scale these approaches.
    
    \item \textbf{Cost Evaluation:} The expenses linked with utilizing the OPENAI API pose a financial challenge that may not be feasible for all organizations. An in-depth cost-benefit analysis of employing sophisticated AI capabilities is needed for a comprehensive assessment.
\end{enumerate}

These challenges needs consideration to advance our methodologies and harness the full potential of LLMs in software development.

\subsection{Future Work}
\label{future Work}
Based on the findings of this research, we plan to investigate several areas in the future:
\begin{enumerate}
    \item \textbf{Bench marking Framework:} Future direction will establish a benchmark  for comparing other open-source LLMs, focusing on content and code generation.
    
    \item \textbf{Exploration of Open Source LLMs:} We aim to assess and integrate open-source LLMs to offer cost-effective options and to perform a comparative analysis of the performance of different models.
    
    \item \textbf{Test Case Code Generation:} The system will be upgraded to produce test case code in user-preferred languages, such as Python or Node.js, meeting specific user needs.
    
    \item \textbf{Co-pilot Development for Testing Tools:} We intend to create a co-pilot feature that automates test suites, specifications, and test case code generation for user requirements, employing advanced AI techniques like RAG and lang-chain on a unified platform.
\end{enumerate}

\section{Conclusions}
\label{conclusions}
In our study, we expand the work of the tool that employs OpenAI, Flask, and React to automatically create test case scenarios from prioritized requirements. This tool represents progress in utilizing AI to enhance test case processes, as it lets users input requirements directly. It showcases the simplification of software testing and documentation by transforming responses into a user-friendly JSON format and providing the capability to download these as CSV files for use with various testing tools.

\bibliographystyle{unsrtnat}
\bibliography{references}  






\end{document}